\documentclass[useAMS,usenatbib]{mn2e}
\usepackage{amsmath}
\usepackage{graphicx}
\usepackage{rotating}
\newcommand{\ba}{\mbox{\boldmath{$\alpha$}}}
\newcommand{\bt}{\mbox{\boldmath{$\theta$}}}
\newcommand{\bb}{\mbox{\boldmath{$\beta$}}}

\def\part#1#2{{\partial #1\over\partial #2}}

\def\A{{\cal A}}

\def\P{{\cal P}}

\begin{document}
\label{firstpage}
\title{The impact of AGN feedback and baryonic cooling on galaxy clusters as gravitational lenses} \author[J. M. G. Mead, L. J. King, D. Sijacki, A. Leonard, E. Puchwein, I. G. McCarthy] {James M. G. Mead$^{1, 2}$\thanks{Email: jmead@ast.cam.ac.uk}, Lindsay
  J. King$^{1, 2}$\thanks{Email:ljk@ast.cam.ac.uk}, Debora Sijacki$^{1,2}$, Adrienne Leonard$^{1,2,3}$ \and Ewald Puchwein$^{4}$ and Ian G. McCarthy$^{2, 1, 5}$  \vspace{0.3cm}\\$^{1}$Institute of Astronomy, University of Cambridge, Madingley Rd, Cambridge CB3 0HA\\
$^2$Kavli Institute for Cosmology, University of Cambridge, Madingley Road, Cambridge CB3 0HA\\
$^3$Service d'Astrophysique, CEA Saclay, 91191 Cif sur Yvette Cedex, France\\
$^4$Max-Planck-Institut f{\"u}r Astrophysik, Karl-Schwarzschild-Stra{\ss}e 1, 85741 Garching, Germany\\
$^5$Astrophysics Group, Cavendish Laboratory, Madingley Road, Cambridge CB3 0HE}
\date{}
\maketitle
\pagerange{\pageref{firstpage}--\pageref{lastpage}} \pubyear{2010}

\begin{abstract} 
We investigate the impact of AGN feedback on the gravitational lensing
properties of a sample of galaxy clusters with masses in the range $10^{14}$ -
$10^{15}\;M_{\odot}$, using state-of-the-art simulations. Adopting a ray-tracing algorithm, we compute the cross-section of giant arcs from clusters simulated with dark-matter only physics (DM), dark matter plus gas with cooling and star formation (CSF), and dark matter plus gas with cooling, star formation and AGN feedback (CSFBH). 
Once AGN feedback is included, baryonic physics boosts the strong lensing cross-section by much less than previously estimated using clusters simulated with only cooling and star formation. For a cluster with a virial mass of $7.4 \times 10^{14}\;M_{\odot}$, inclusion of baryonic physics without feedback can boost the cross-section by as much as a factor of 3, in agreement with previous studies, whereas once AGN feedback is included this maximal figure falls to a factor of 2 at most. Typically, clusters simulated with DM and CSFBH physics have similar cross-sections for the production of giant arcs. We also investigate how baryonic physics affects the weak lensing properties of the simulated clusters by fitting NFW profiles to synthetic weak lensing data sets using a Markov Chain Monte Carlo approach, and by performing non-parametric mass reconstructions. Without the inclusion of AGN feedback, measured concentration parameters can be much larger than those obtained with AGN feedback, which are similar to the dark-matter only case.

\end{abstract}


\section{Introduction}
In recent years, strong gravitational lensing by galaxy clusters has been used to investigate cosmological problems and to improve our understanding of cluster physics. Strong lensing can manifest itself as highly non-linear distortions of light from background galaxies (giant arcs) and observational searches have revealed many spectacular examples around both optically-selected and X-ray-selected clusters (e.g. \cite{Zaritsky}, \cite{gladders}).  It has been found that the strong lensing properties of clusters depend on myriad physics - including the cosmological parameters, halo substructure and baryonic physics. Strong lensing is hence a good technique to investigate both cosmology and cluster physics. 

The frequency of giant arcs around galaxy clusters is sensitive to the cosmological model. Using dark matter simulations, \cite{BHC} (henceforth B98) found that in $\Lambda$CDM the number of giant arcs 
is a factor of 5-10 lower than in observations. This disparity arises either because our cosmological model is incorrect or, more likely, that some aspects of the modeling of the source or lens populations are incorrect, for example that we are missing crucial physics from our cluster simulations. 

A substantial body of work has investigated this latter point, using increasingly sophisticated simulations in order to reconcile predictions with observations. There have been a number of papers investigating the strong-lensing properties of dark matter-only clusters (e.g. \cite{HoW}, \cite{Hilbert}). The earliest analytic work focused on spherically symmetric cluster models, but it has been demonstrated that halo triaxiality has a strong impact on arc production e.g. by \cite{Oguri}, who used semi-analytic models, and in a complementary analysis by \cite{Dalal} who analysed a simulated cluster sample.

Other papers have also considered the impact of baryonic physics, in particular models including gas with cooling and star formation, and cluster galaxies. Baryons can have a significant impact on the central regions of galaxy clusters, and since it is precisely this region that is important in arc production, one would expect the abundance of giant arcs to be sensitive to the baryonic physics of the cluster. \cite{Puchwein}, \cite{WamOs} and \cite{Rozo} found that the addition of baryons to cluster simulations increases the lensing cross-section by factors of a few. Of the cluster galaxies, only the central cD galaxy makes a non-negligible contribution to the strong lensing properties, increasing cross-section by up to 50\% \citep{Men03}. 

The baryonic cluster simulations which have been used to date in the prediction of lensing cross-sections suffer from significant over-cooling, with much higher stellar fractions than observed (see the discussion in e.g. \cite{PSS}, \cite{Balogh} and \cite{PSSK}). In the absence of heating, the central regions of clusters cool faster than the outer regions, causing gas to flow inward in a quasi-hydrostatic cooling flow; this would result in a large quantity of cold gas in the central regions of clusters, a high star formation rate, and extremely luminous central galaxies. However, this is at odds with both observations of clusters, which do not show such strong cooling flows, and also of Brightest Cluster Galaxies (BCGs) that tend to have evolved, red, stellar populations. There is compelling evidence from observational and theoretical work that the AGN feedback processes that accompany the accretion of gas onto supermassive black holes - residing in most galaxies - provide a heating mechanism that prevents over-cooling, bringing physical models of clusters more into line with observations (e.g. \cite{ValSilk}; \cite{Bower}; \cite{McNNul}; \cite{McCarthy}; \cite{Croton}). 

It is well established that strong lensing cross-sections can be used to probe the central density profiles of clusters - steeper slopes of inner density profiles enhance arc production capability, e.g. \cite{Men07}. \cite{Torri} investigated the effects of cluster mergers on the strong lensing cross-section, finding that during a merger the cross-section can be amplified by up to an order of magnitude. The cross-section for strong lensing can also be boosted by as much as 50\% for lower mass clusters by line-of-sight structure \citep{PuchHil}. An additional factor that can affect the cross-section are the properties of the background galaxies themselves, such as their redshift distribution, size, shape and clustering, e.g. \cite{Wam04}; \cite{Li}; \cite{Gao}. 

Departures from $\Lambda$CDM have also been invoked to reconcile the observed abundance of arcs with the value of $\sigma_{8}$ ($\sim 0.8$) consistent with cosmic shear surveys and CMB observations from WMAP. For example, \cite{BDW} consider early dark energy (EDE) cosmologies, where there is significant dark energy even at very early epochs, unlike the case of $\Lambda$CDM or many simple quintessence models. Their analytic predictions suggest that non-linear structure formation can be substantially enhanced beyond that of $\Lambda$CDM. \cite{Fedeli} suggest that EDE has a significant impact on the statistics of giant arcs, bringing predictions and observations into closer agreement. More recently, \cite{Francis}  and \cite{GrossiSpringel} carried out the first studies of non-linear structure formation in EDE models using N-body simulations, finding its impact to be much more difficult to detect observationally than implied by analytic models. 

There has also been interest in the effects of baryonic physics on the weak lensing properties of clusters.
Future large surveys, such as the Dark Energy Survey (DES; https://www.darkenergysurvey.org/) will measure the matter power spectrum with unprecedented precision, also allowing stringent constraints to be placed on the properties of dark energy \citep{Albrecht}. Studies by, for example, \cite{Rudd} and \cite{White}, have indicated that the inclusion of baryonic physics can have a significant effect on the weak lensing properties of clusters, and hence would significantly alter the predictions for the matter power spectrum relative to those simulations which include only dark matter. 

It is essential to understand the impact that baryonic physics and AGN
feedback might have on the gravitational lensing properties of galaxy
clusters in order to calibrate constraints on their physical properties, and fully exploit their potential as cosmological probes. This is important
both for the interpretation of existing data, and to prepare for upcoming
large surveys such as DES. In this paper we seek to extend the picture of
cluster lensing by investigating the influence of AGN feedback on the lensing
properties of numerically simulated clusters. We use recent state-of-the-art
simulations from \cite{PSS} that incorporate the growth of black holes (BHs)
and corresponding feedback processes using the prescription detailed in
\cite{Sijacki}. These are the first cosmological simulations of clusters to
incorporate AGN feedback self-consistently. We use these simulations to demonstrate the impact that 
realistic modelling of AGN physics can have on cluster lensing results.

Throughout this paper a $\Lambda$CDM cosmology with parameters $\Omega_{\rm m} = 0.25$, 
$\Omega_{\Lambda} = 0.75$ and $H_{0} = 73\, \rm{km\,s^{-1} Mpc^{-1}}$ is adopted. The paper is organized as follows. In Section 2 the cluster simulations that are used are described. The numerical methods for the creation of synthetic arcs using strong lensing ray-tracing through the simulated clusters, 
and the generation of weak lensing catalogues are outlined in Section 3. The impact of baryonic physics on the strong-lensing and weak-lensing properties of the clusters are presented in Section 4 and in Section 5 respectively. Finally, in Section 6 the significance of our findings in the context of previous work is discussed.

\section{The Cluster Simulations}
\label{model}
We consider the gravitational lensing properties of five galaxy
clusters at $z=0.2$ with virial masses in the range
$1.5\times10^{14}-7.4 \times 10^{14}\,M_{\odot}$, each with three types of physics:
dark-matter only (DM), dark matter with gas, cooling and star formation (CSF)
and the latter with added AGN feedback (CSFBH). The mass range was chosen to
be large enough to investigate any strong mass dependence of lensing
properties on baryonic physics. The virial masses and radii of these clusters
are summarised in Table \ref{clustertable}. The virial
mass and radius are taken to be equivalent to $M_{200}$ and $r_{200}$, defined
by the mass enclosed inside the radius within which the mass density is 200
times that of the critical density of the universe at the redshift of the
cluster. Note that in Table 1, clusters are designated C1-C5 in order of increasing 
$M_{200}$ with respect to the background density. The cluster sample is taken 
from \cite{PSS}, in which the simulations
are based on the numerical prescription described in
\cite{Sijacki}. \cite{PSS} selected clusters from the dark-matter-only
Millennium simulation \citep{SpringNat}, and performed a number of
re-simulations that incorporated baryonic physics with and without AGN
feedback. DM clusters that are used in this paper were not already available
at as high a resolution as the CSF and CSFBH runs, so these were re-simulated
using the same process as in \cite{PSS} (summarised in Section
\ref{resim}). In Section \ref{code} we recap on the numerical code, and in
Section \ref{BHmodel} the black hole model is discussed in more depth. 

\begin{table}
\caption{The virial masses and radii (physical units) of the CSFBH clusters used in
  this paper at $z=0.2$.}
\centering
\begin{tabular} {c c c}
\hline \hline
 Cluster & $M_{200}$ & $r_{200}$ \\
& $(10^{14}\,M_{\odot})$ & (Mpc) \\ [0.5ex]
\hline
C1 & 1.5 & 0.88\\
C2 & 2.1 & 1.00\\ 
C3 & 4.2 & 1.25\\
C4 & 3.1 & 1.13\\
C5 & 7.4 & 1.50\\ [1ex]
\hline
\end{tabular}
\label{clustertable}
\end{table}

\subsection{Re-simulating Millennium Clusters}
\label{resim}
The re-simulations were performed by selecting the Lagrangian region of the target halo and then populating it with a larger number of lower mass particles, adding small-scale power up to the new Nyquist frequency. At larger distances outside this high-resolution region, increasingly more massive particles were used, ensuring that the gravitational tidal field acting on the high resolution region is accurately represented. For the simulations with gas, each high-resolution dark matter particle is split into a dark matter particle and a gas particle, displacing them by half of the original mean inter-particle separation. The centre of mass of each pair is fixed. 

\subsection{The Numerical Code}
\label{code}
The simulation of the cluster sample in \cite{PSS} follows the same
prescription as in \cite{Sijacki} and readers are referred to this paper for
more details on the simulation and black hole (BH) model.  The Tree-SPH code
{\small GADGET-3} is used (based on \cite{Springel2005}), which includes gravity and
non-radiative hydrodynamics of dark matter and gas components, and also
follows radiative cooling and heating of an optically thin plasma of hydrogen
and helium. A sub-resolution multiphase model for star formation and the
associated supernova feedback was adopted \citep{SpHern}. The growth and
feedback processes of a population of BHs embedded in the simulations were
self-consistently followed, and this process is described in section
\ref{BHmodel}.

\subsection{The Black Hole Model}
\label{BHmodel}
In the simulations, BHs are represented as collisionless sink particles which can accrete gas from their surroundings. Any halo above a threshold mass of $5 \times 10^{10}\,h^{-1}\,M_{\odot}$ is seeded with a BH of mass $10^5\,h^{-1}\,M_{\odot}$, if it does not contain a BH already. The growth of this BH population through accretion is then followed. The accretion rate of the BH particles follows the spherically symmetric Bondi-Hoyle-Lyttleton model (\cite{HoyLyt}; \cite{BondHoy}; \cite{Bondi}):

\begin{equation}
\dot{M}_{\rm{BH}} = \frac{4 \pi \alpha G^{2} M^{2}_{\rm{BH}} \rho}{\left( c^{2}_{s} + v^{2} \right)^{3/2}}\;,
\end{equation}

\noindent where $\alpha$ is a dimensionless parameter ($\alpha=100$ in the simulations
we use), $\rho$ is the density, $c_{s}$ is the sound speed of the gas and $v$
is the velocity of the BH relative to the gas. In the model, the upper limit
of the BH accretion rate is given by the Eddington rate.

\begin{figure*}
\centering
\begin{tabular}{ccc}
\includegraphics[width=2.2in]{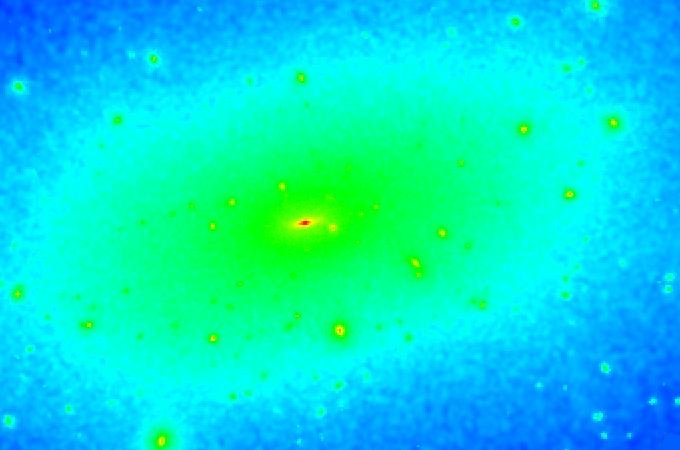}&
\includegraphics[width=2.2in]{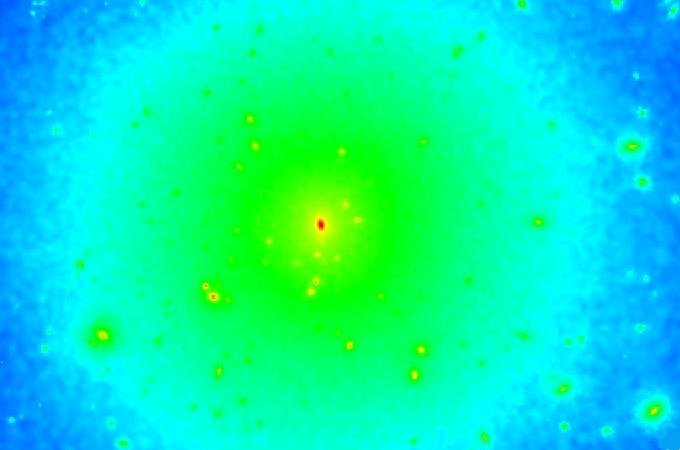}&
\includegraphics[width=2.47in]{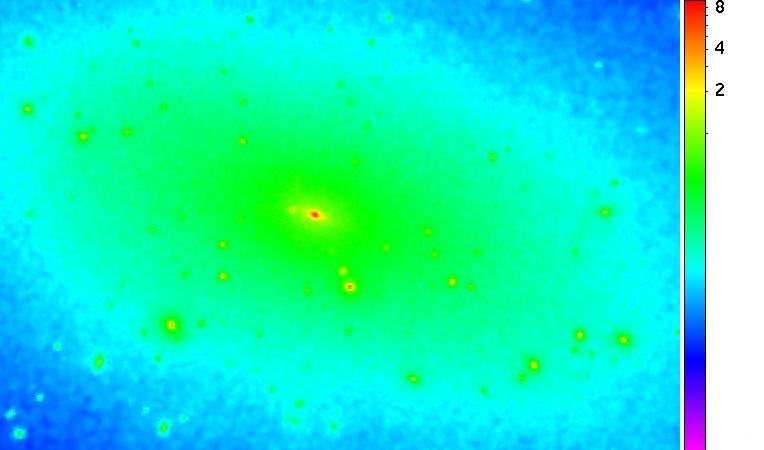} \\
\includegraphics[width=2.2in]{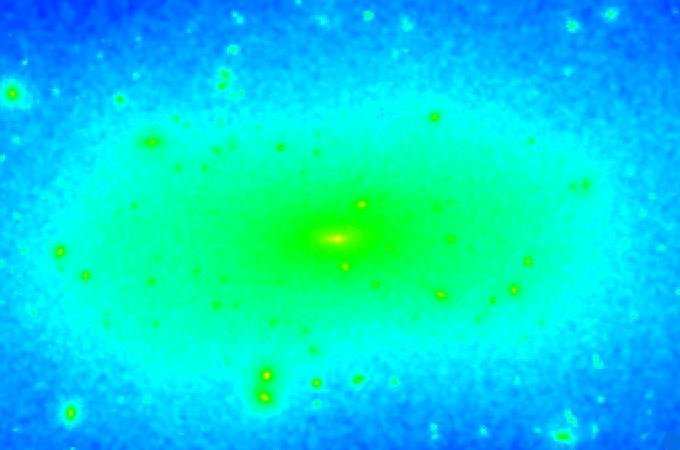}&
\includegraphics[width=2.2in]{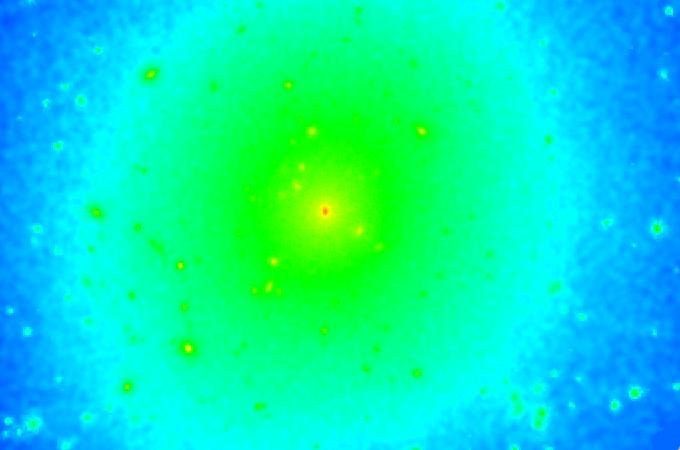}&
\includegraphics[width=2.47in]{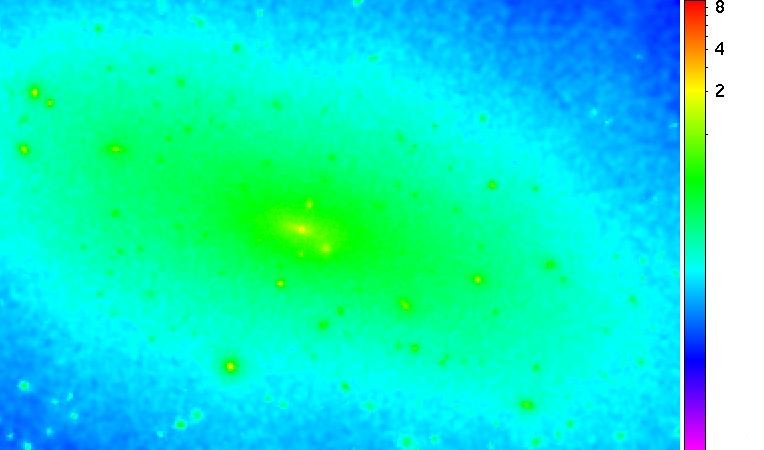} \\
\includegraphics[width=2.2in]{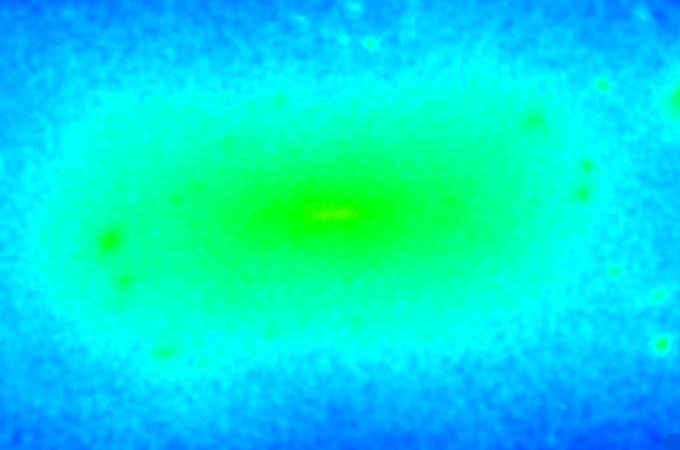}&
\includegraphics[width=2.2in]{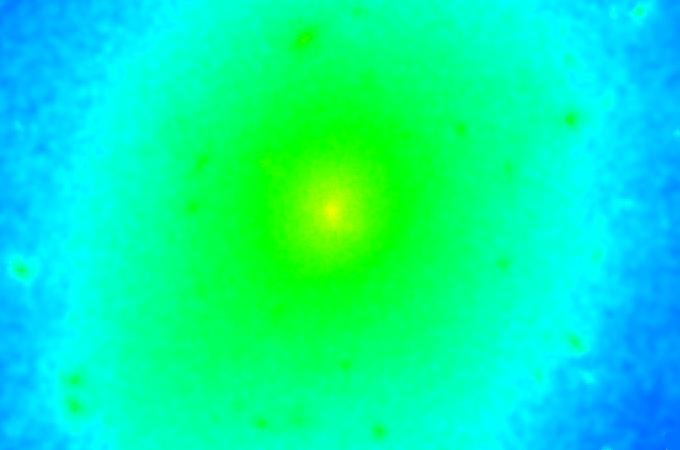}&
\includegraphics[width=2.47in]{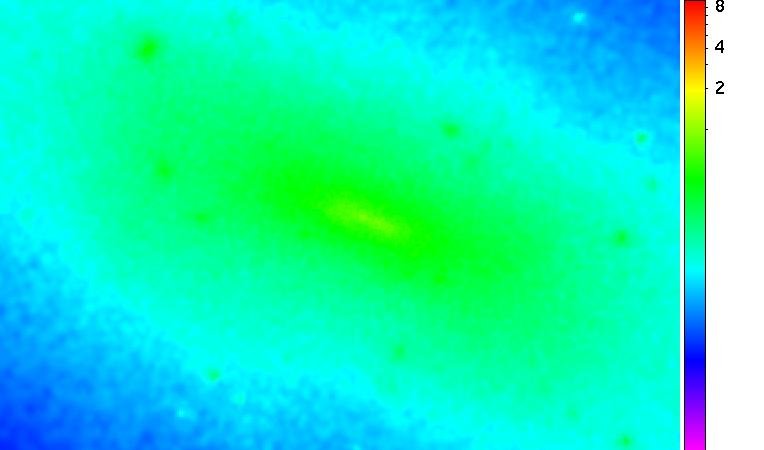} \\
\end{tabular}
\caption{Maps of the convergence for cluster C5 using CSF (top row), CSFBH (middle row) and DM (bottom row) for sources at $z=3$. We show 3 different projections for each physical prescription, reading left to right - xy, yz and xz. The dimensions of the maps are 600\,kpc in height and 900\,kpc in length.The scale bars on the right indicate the ranges of convergence.}
\label{kappamaps}
\end{figure*}

There is one additional process through which BHs may gain mass - they are
allowed to merge with other BHs that lie within the smoothing lengths  used to
estimate the local gas density, and have relative velocities lower than the
local gas sound speed.

The model for BH feedback accounts for AGN heating at both high and low
accretion rates. In the model there are two modes of AGN feedback: at high
accretion rates it is assumed that the bulk of AGN heating originates in
luminous quasar activity, whilst at lower accretion rates, corresponding
typically to lower redshifts, AGN heating proceeds via radiatively inefficient
feedback in a mostly mechanical form. The transition between these two regimes
is characterized by a parameter $\chi_{\rm{radio}} =
\dot{M}_{\rm{BH}}/\dot{M}_{\rm{Edd}}$ - for $\chi_{\rm{radio}} > 10^{-2}$ we
are in the ``quasar" heating regime, and for $\chi_{\rm{radio}} \le 10^{-2}$ we
are in the``radio mode" feedback regime, which is modelled by injecting
bubbles into the host cluster.

In the quasar heating regime a small fraction of the bolometric luminosity is
coupled thermally and isotropically to the surrounding gas particles
(\cite{Spring05}), with an amount given by,

\begin{equation}
\dot{E}_{\rm{feed}} = \epsilon_{\rm{f}}L_{\rm{r}} = \epsilon_{\rm{f}} \epsilon_{\rm{r}} \dot{M}_{\rm{BH}} c^{2}\;.
\end{equation}

In this equation $\epsilon_{f}$ is the efficiency of thermal coupling and
$\epsilon_{r}$ is the radiative efficiency. We adopt $\epsilon_{f} = 0.05$ and
$\epsilon_{r} = 0.1$ as in previous work, to ensure agreement between the
simulated and observed $M_{\rm{BH}} - \sigma_{*}$ relation (\cite{Sijacki},
\cite{DiMatteo}).

In the radio mode feedback regime, below $\chi_{\rm{radio}}$, accretion
intermittently produces AGN jets which blow hot bubbles in the surrounding
gas. The cluster simulations that we use employ a model of radio feedback in
which an AGN-driven bubble is created if a BH has increased in mass by a
fraction $\delta_{\rm{BH}} \equiv \delta M_{\rm{BH}}/M_{\rm{BH}}$. In this
case, the energy content of the bubble is,

\begin{equation}
E_{\rm{bub}} = \epsilon_{\rm{m}} \epsilon_{\rm{r}} c^{2} \delta M_{\rm{BH}}\;,
\end{equation}

where $\epsilon_{\rm{m}} = 0.2$ is the mechanical feedback efficiency. The
radius of the bubble is modelled by,

\begin{equation}
R_{\rm{bub}} = R_{\rm{bub,0}}\left(\frac{E_{\rm{bub}}/E_{\rm{bub,0}}}{\rho_{\rm{ICM}}/\rho_{\rm{ICM,0}}} \right)^{1/5}\;,
\end{equation}

where $R_{\rm{bub,0}}$, $E_{\rm{bub,0}}$ and $\rho_{\rm{ICM,0}}$ are
normalization constants for bubble radius, energy content and ambient density.

\section{Numerical Methods for Lensing Simulations}

Surface mass density maps are created by projecting the matter on to a
two-dimensional grid using an adaptive SPH projection algorithm. It is also
possible to project using a triangular-shaped clouds (TSC) scheme. Where the
particle density is high (i.e. in the strong-lensing region around the centre
of the cluster) the two algorithms should produce almost identical surface
densities. We verified that the choice of projection algorithm does not affect
the lensing cross-sections. 

The physical extents of the clusters are much less than the distances between the observer-cluster and cluster-sources, so the thin-lens approximation is valid. Each surface mass density map is determined on a 600 $\times$ 600 grid covering a square of 1.5$\,h^{-1}$\,Mpc (comoving) on a side, centred on the cluster. To create a map of the convergence ($\kappa$), the surface density map, $\Sigma$, is divided by the critical surface mass density, 
\begin{equation}
\Sigma_c=\frac{c^2}{4\pi G}\frac{D_{\rm s}}{D_{\rm ds}D_{\rm d}}\;,
\end{equation}
where $D_{\rm s}$, $D_{\rm d}$ and $D_{\rm ds}$ are the observer-source, observer-lens and lens-source angular diameter distances.  In Fig.\,\ref{kappamaps} we show the convergence for cluster C5, with all 3 types of physics (CSF, DM, CSFBH) in 3 projections, for sources at $z=3$.

\subsection{Ray-tracing for Strong Lensing Arc Production}
The strong lensing properties of a cluster are sensitive to the mass
distribution in the inner regions. We now describe how synthetic maps of
distant galaxies strongly lensed by simulated clusters are obtained.  The
ray-tracing for the simulation of strong lensing by a cluster, that follows
light rays between the source and observer via deflection at the cluster, is
performed by an algorithm used in \cite{LKW}. The mapping between the source
and lens (image) planes is given by the `lens equation',

\begin{equation}
\bb = \bt -\ba \left( \bt \right)\;,
\end{equation}
where $\bb$ are the coordinates in the source plane, $\bt$ are the coordinates in the lens plane, and $\ba$ are the deflection angles. The lensing deflection potential $\psi \left(\bt \right)$ is related to $\kappa$ through a Poisson equation,
\begin{equation}
\label{poiss}
\nabla^{2} \psi = 2 \kappa\,.
\end{equation}

\noindent We can also relate $\psi \left(\bt \right)$ to $\ba$ via,
\begin{equation}
\nabla \psi = \ba\,.
\end{equation}
Therefore it follows that $\kappa$ can be related to $\ba$ in Fourier space,
\begin{equation}
\tilde{\alpha}_{i} = -\frac{2ik_{i}}{k_{1}^{2} + k_{2}^{2}} \tilde{\kappa}\;.
\label{raytrace}
\end{equation}

To simulate strong lensing, equation \ref{raytrace} is used to calculate the deflection angles due to the cluster (on a grid of twice the resolution of the input convergence map). To achieve greater numerical accuracy, and to mitigate the influence of edge effects, the convergence field is zero-padded. The output of this ray-tracing program is a map containing the lensed arcs. This map is searched for arcs using the algorithm presented in \cite{Horesh}. The algorithm makes calls to SExtractor \citep{SE}, using slightly different detection parameters in each call, to detect objects with an axis ratio above a given value. The arc length-to-width ratio is measured as in \cite{Miralda}.

The giant arc cross-sections are computed using a Monte Carlo approach. The cluster field is populated with background galaxies placed at randomly assigned positions, with exponential profiles and elliptical shaped isophotes drawn as Gaussian random deviates, as described in \cite{LKW}. This galaxy population is placed at uniform redshift, $z_{\rm g} = 3$ - although the redshift distribution of the source galaxy population is important for determining the absolute cross-section, in this work we are interested in the relative cross-sections of clusters simulated with different types of physics. This population of galaxies is then ray-traced through the simulation, using the methods described above. In this work, we focus on arcs with a length-to-width ratio $r \geq 7.5$, making no distinction between tangential and radial arcs. To calculate the cross-section, 100 populations of background galaxies are simulated, each with a density of 50 arcmin$^{-2}$, and the number of arcs produced in each realisation is recorded. The cross-section is then computed by multiplying the fraction of sources that form giant arcs by the area of the cluster field over which those galaxies were distributed. An error on the cross-section is estimated by repeating this procedure 10 times (in all we simulate 1000 populations behind each cluster), and computing the standard deviations of the cross-sections.

\subsection{Weak Lensing Simulations}
Weak lensing is sensitive to the shear field of the cluster on larger scales. The weak lensing simulations that are undertaken to generate synthetic catalogues of weakly lensed distant galaxies behind the simulated clusters are now described. These catalogues are used to obtain NFW fits to the total mass profile, and non-parametric mass reconstructions of the target clusters.

\begin{figure*}
\centering
\begin{tabular}{cc}
\includegraphics[width=3.5in]{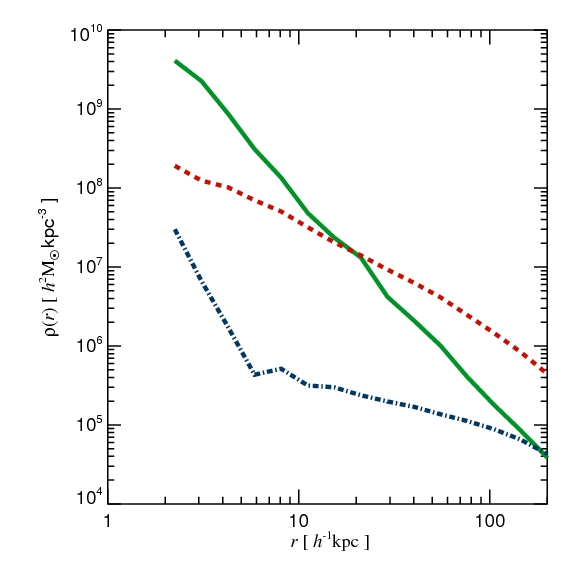}&
\includegraphics[width=3.5in]{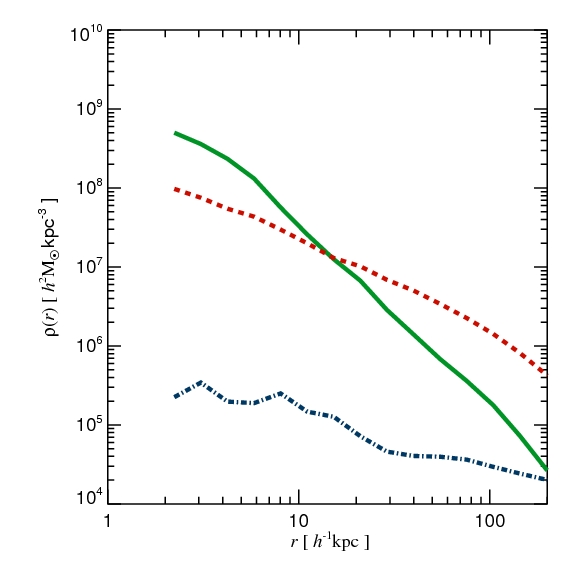}
\end{tabular}
\caption{Spherically averaged density profiles of cluster C4. The left-hand
  panel shows the cluster simulated with cooling and star formation (CSF) and
  the right-hand panel shows the same cluster simulated with AGN feedback included (CSFBH). The solid green curve represents stellar density, the red dashed curve represents the density of dark matter and the dot-dashed blue curve shows the density of gas.}
\label{clusterprofile}
\end{figure*}

Whereas the convergence magnifies background objects such as galaxies, the shear stretches them. The complex shear (a complex quantity with magnitude corresponding to the shear amplitude, and phase equal to twice the shear position angle) also depends on the second derivatives of the lensing potential,
\begin{equation}
\label{gamma}
\gamma = \gamma_{1} + i \gamma_{2} = \left(\psi_{,11} - \psi_{,22} \right)/2 + i \psi_{,12}\;,
\end{equation}
where commas denote partial differentiation. The complex reduced shear is obtained from the shear and convergence through the equation $g=\gamma/(1-\kappa)$. Using the $\kappa$ maps corresponding to clusters simulated with DM, CSF or CSFBH physics, the corresponding maps of $\gamma$ and $g$ are obtained, by relating $\kappa$ and $\gamma$ in Fourier space given Eq.\,\ref{poiss} and Eq.\,\ref{gamma}.

The magnification is given by the inverse Jacobian determinant of the lens equation, $\mu(\bt) = [\det \A \left( \bt \right)]^{-1}$. Evaluating this we obtain,
\begin{equation}
\mu = \frac{1}{\left[\left(1-\kappa \right)^{2} - \gamma^{2} \right]}\;.
\label{emu}
\end{equation}

The ellipticity of a galaxy, describing its shape and orientation, is denoted by a complex number with a modulus related to the axis ratio $r$ of $(1-r)/(1+r)$, and phase being twice the position angle.
In the non-critical regime ($\det \A > 0$) the weakly lensed image of a distant galaxy has a complex ellipticity given by,
\begin{equation}
\label{WL}
\epsilon=\frac{\epsilon^{\rm s}+g}{1+g^{*}\epsilon^{\rm s}}\;,
\end{equation}
\noindent where $\epsilon$ and $\epsilon^{\rm s}$ are the lensed and unlensed complex ellipticities respectively, and $g^{*}$ is the complex conjugate of the reduced shear \citep{BSrev}. 

Galaxy ellipticities are taken to have a Gaussian probability density function \citep{BSrev}, 
\begin{equation}
\label{epdf}
p_{\epsilon^{\rm s}}= \frac{\exp\left({-|\epsilon^{\rm s}|^{2}/\sigma_{\epsilon^{\rm s}}^{2}}\right)}{\pi\sigma^{2}_{\epsilon^{\rm s}}\left[1-\exp\left({-1/\sigma_{\epsilon^{\rm s}}^2}\right)\right]}\;.
\end{equation}
Throughout this work we take $\sigma_{\epsilon} = 0.2$. 

In the weak-lensing analysis, the simulated background galaxies are at $z_{\rm g} = 3.0$, with random positions on the sky, and with ellipticities drawn from the Gaussian probability density function given by Eq.\,\ref{epdf}. The source galaxy redshift was chosen to be sufficiently high so as to ensure that the low mass clusters would have cross-sections significantly above zero. These galaxies are then lensed by the cluster according to Eq.\,\ref{WL}, using the map of $g$ corresponding to the $\kappa$ map for a particular cluster, and taking into account any depletion due to magnification. The slope of the unlensed number counts is taken to be $s=0.5$, so that the local lensed number density $n_{\rm L}$ is related to the local unlensed number density $n$ via $n_{\rm L}=n\,\mu^{s-1}$. We use a number density of 150 arcmin$^{-2}$, well within the power of future surveys \citep{lewking}, although out of reach of present observing capabilities. With future observations we ought to be able to probe the differences in the weak-lensing properties of clusters in great detail.

\section{Strong Lensing Results}

In this Section our key strong-lensing result, the effect of AGN feedback on strong lensing cross-sections, is presented. In Fig.\,\ref{clusterprofile} we show the spherically averaged density profiles of stars, gas 
and dark matter in cluster C4, simulated both with CSF and CSFBH. The stellar density is much higher in the CSF run in the central regions of the cluster - a result of the `overcooling problem'. Also note that the dark matter density profile has contracted in the CSF simulation relative to CSFBH, due to the comparatively large density of baryons in the cluster centre (\cite{Blumenthal}, \cite{Gnedin}).

\begin{figure*}
\begin{tabular}{lll}
\hspace*{-0.3cm}
\includegraphics[width=2.3in]{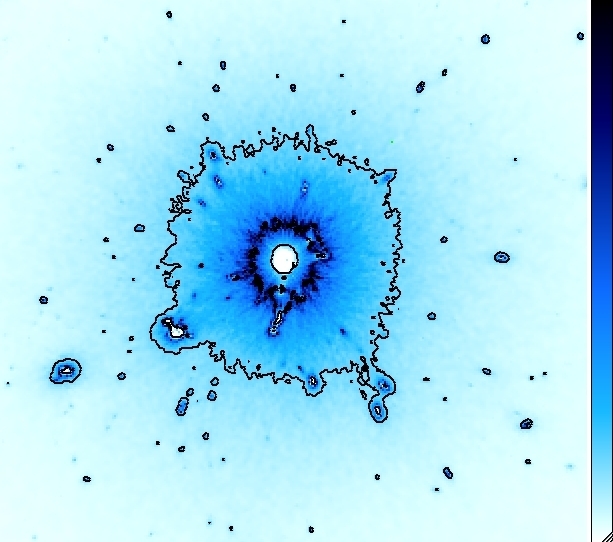}&
\includegraphics[width=2.3in]{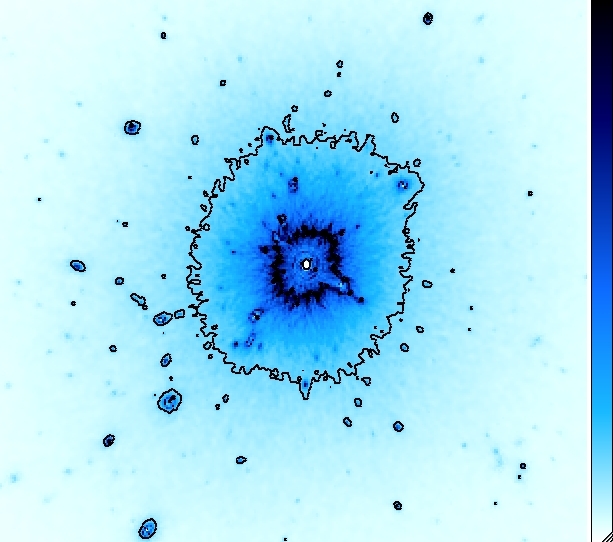}&
\includegraphics[width=2.3in]{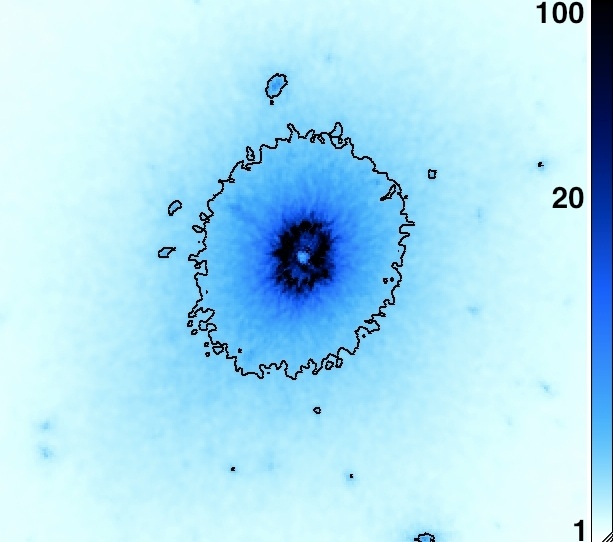}
\end{tabular}
\caption{Maps of the magnification for sources at $z=3$ in the inner 0.75$\,h^{-1}$\,Mpc of the most massive cluster C5 with CSF, CSFBH and DM physics from left to right respectively. The magnification is shown on a log scale, with the values of the scale indicated on the right panel. The black contour corresponds to magnification of 2.}
\label{clumag}
\end{figure*}

Fig.\,\ref{clumag} shows the image plane magnification for sources at $z_{\rm g}=3$ in the inner 0.75$\,h^{-1}$\,Mpc  of the most massive cluster C5 with CSF, CSFBH and DM physics, calculated using Eqn.\,\ref{emu} from the convergence and shear maps. The observed properties of strongly lensed arcs depend on the distribution of magnification.

 \subsection{The impact  of baryonic physics on cross-section}

\begin{figure*}
\centering
\begin{tabular}{ll}
\hspace*{-0.5cm}
\includegraphics[scale=0.7]{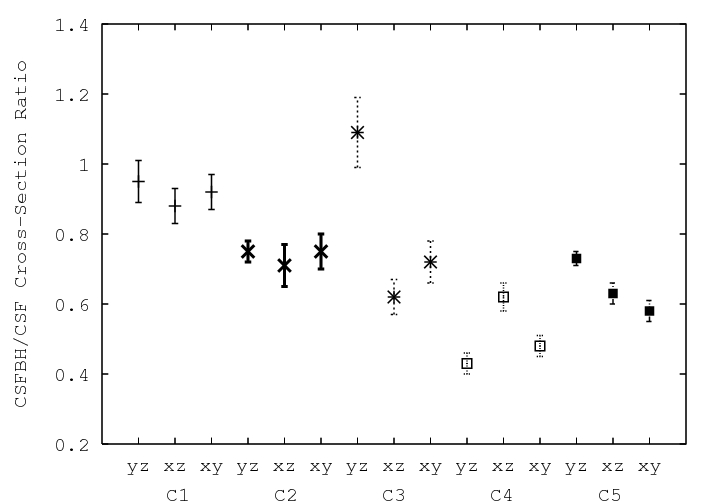} &
\includegraphics[scale=0.7]{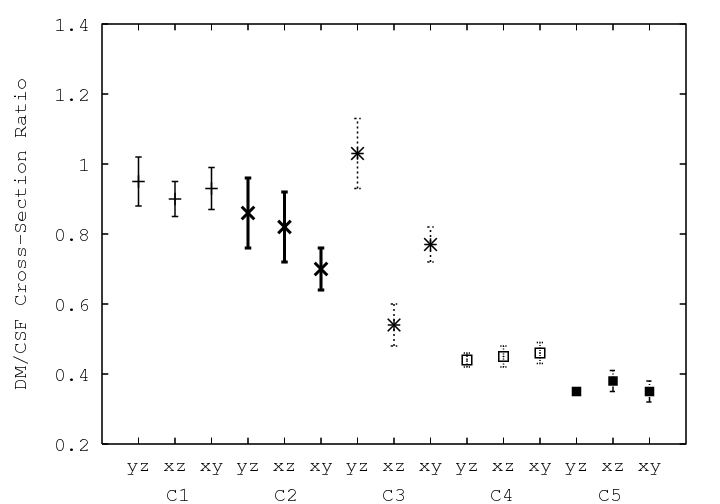}\\\\
\hspace*{-0.5cm}
\includegraphics[scale=0.7]{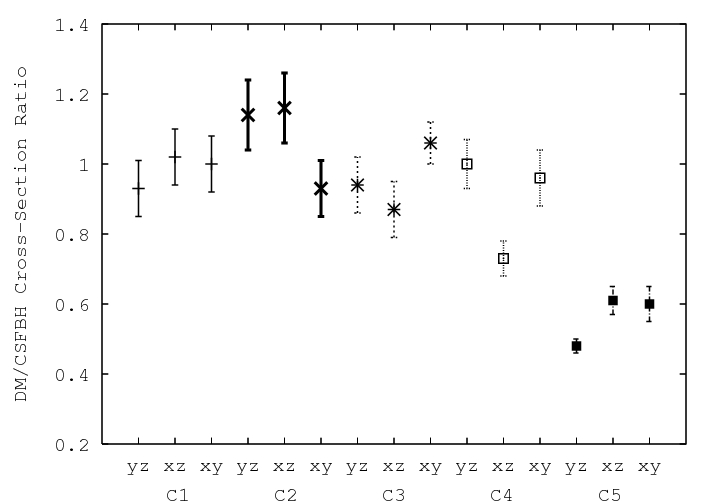}& 
\end{tabular}
\caption{Ratios of strong lensing cross-sections for the 3 orthogonal projections of each cluster. 
In Fig.\,4a, 4b and 4c we show the ratio of cross-sections CSFBH/CSF, DM/CSF and DM/CSFBH respectively. On the x-axis, the xy, yz and xz projections of C1 through C5 are denoted (cycling through the projections from the lowest to highest $M_{200}$ with respect to the background density).}
\label{ratios}
\end{figure*}

The ratios of the cross-sections were calculated for DM, CSF and CSFBH clusters and the results are summarised in Fig \ref{ratios}. 
Using a source plane definition for arc cross-section, \cite{Rozo}
and \cite{Puchwein} found that the inclusion of baryonic cooling boosted
cross-sections of the most massive clusters by factors of approximately 2-4
above those of DM clusters. We can see that the results of this paper are
broadly in agreement with this result - for instance the CSF cross-sections of
C5 exceed those of the DM simulation by factors of 2-3
(Fig.\,\ref{ratios}). However, on inclusion of AGN feedback we see that for
all but cluster C5, the CSFBH and DM cross-sections are comparable. Even for cluster C5, the boost in cross-section above the DM cluster
gained by adding baryons tempered with AGN feedback is $< 2$. For all clusters the reduction in cross-section between CSF and CSFBH is statistically significant. For clusters C2, C4 and C5 clearly the cross section has been reduced by many times the typical error bar. The yz- projection of C3 is anomalous insofar as addition of AGN physics caused no notable reduction in cross-section. On closer analysis, C3 appears to be a merging cluster with large amounts of substructure close to the cluster centre - in this sense, this projection of the cluster is `atypical'. The results of C1 perhaps warrant closer statistical analysis; for this purpose, we can use Student's t-test with the null hypothesis that the mean cross-section of the two samples of clusters (CSF and CSFBH) are equal. It was found that in all cases we could reject the null hypothesis with greater than 95\% confidence. 

The results can be explained by analyzing the density profiles of the
clusters. \cite{Rozo}, amongst others, have demonstrated that haloes with
steeper inner density profiles have higher strong lensing
cross-sections. Thus, the addition of baryons to clusters is expected to boost
strong-lensing cross-sections, as gas condenses in the centres of these
haloes, and so the density profiles are steepened significantly relative to
the DM-only case (e.g. Fig.\,\ref{clusterprofile}). In the same way, because
the CSF haloes have a greater central concentration of baryons than the CSFBH
haloes, the central density profile is steeper for a CSF cluster than for a
CSFBH cluster. It is thus unsurprising that the CSFBH haloes prove less
efficient strong lenses than CSF haloes. The redshift of background galaxies will also have an effect on the results. For background galaxies with a relatively high redshift, poor clusters with CSF gas physics could have a projected surface mass density greater than the critical density, whereas either (or both) of DM or CSFBH clusters could remain sub-critical. This would cause a bias. The results of this paper will actually be strengthened through use of a lower source redshift - CSF clusters, due to their extremely high central density, will almost always exceed critical density; it is CSFBH and DM clusters that could first drop to sub-critical levels when lower source redshifts are used - this would result in an even greater difference between CSF and CSFBH cross-sections. To check this result, the cross-sections of C4 were investigated using a background population at $z_{g} = 1$. In all cases the ratio between CSFBH and CSF cross-sections was reduced relative to the $z_{g} = 3$ case; for example, for the xy- projection, the new ratio was calculated as 0.52. As described above, the reduction will be even more pronounced for lower mass clusters.

\begin{figure}
\centering
\begin{tabular}{cc}
\begin{sideways} ~~~~~~~~~~~~CSF/CSFBH Density Ratio\end{sideways} &
\hspace{-2cm}
\includegraphics[width=3.5in]{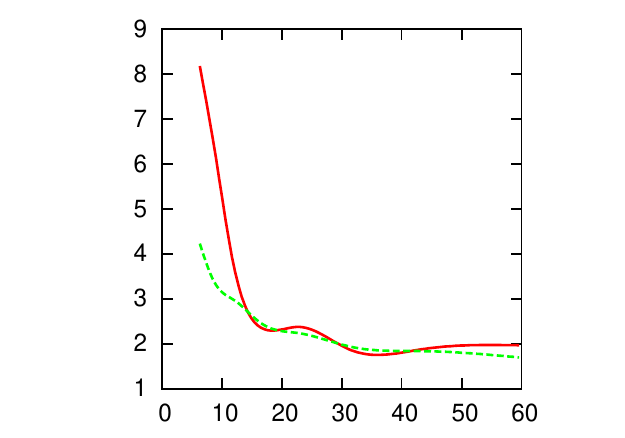}\\
& \hspace{-1cm}r [kpc]
\end{tabular}
\caption{The density profile of the CSF simulation divided by the density profile of the CSFBH simulation for clusters C2 and C4. The result is shown for the inner 60 kpc, as it is the density profile in this region that primarily affects strong lensing cross-sections. The solid line represents C4 and the dashed line corresponds to C2.}
\label{rhoratio}
\end{figure}

In order to illustrate the relative effect of AGN feedback for two different clusters, in Fig.\,\ref{rhoratio} the density profile of the CSF simulation divided by the corresponding CSFBH density profile is plotted for clusters C2 and C4. In C4, the density profile is affected comparatively more by the inclusion of AGN feedback, both in terms of its absolute value and its slope.

As regards to the mass trend, a much larger sample will be required to draw firm
conclusions. With the exception of C5 (which is discussed below), with increasing cluster mass the ratio CSFBH/CSF decreases, whilst the ratio DM/CSFBH is approximately constant. We can physically explain these mass trends as follows. The AGN feedback is less effective in cluster C5 than C4, insofar as the CSFBH cross-section is a larger fraction of the CSF cross-section, and
exceeds the DM cross-section by a greater amount. A possible explanation lies
in the fact that AGN feedback does not completely overcome central gas
  cooling in C5. Thus, relatively speaking, the AGN feedback has less of an impact in C5 than in C4. The stellar fraction of C5, even when simulated with CSFBH, is still larger than observed in reality \citep{PSS} - if the stellar fraction were reduced in C5 to observed levels this would almost certainly result in a further reduction in the CSFBH cross-section, leading both to a reduction in the ratio CSFBH/CSF and an increase in the ratio DM/CSFBH. On looking at Fig. \, \ref{ratios} we find that by reducing the ratio CSFBH/CSF and increasing DM/CSFBH we could possibly fit C5 to both postulated trends - that the ratio CSFBH/CSF falls with increasing cluster mass, and the ratio DM/CSFBH is approximately constant. On the other hand, the AGN feedback is more effective in cluster C4 than in C2 - in C4 the CSFBH cross-section is a smaller fraction of the CSF cross-section, whilst in both cases the DM and CSFBH cross-sections are comparable. This can be explained by looking at Fig.\,\ref{rhoratio} - relatively speaking, the AGN feedback has had a greater effect on both the magnitude and slope of the central density profile in cluster C4 as opposed to cluster C2. 

It is well known that for an individual cluster the absolute cross-section of a cluster will vary significantly depending on its projection, and we observe this in our clusters. It is also worth pointing out that although a mass trend does appear to be observed in our results, effects of triaxiality and asymmetry effects in clusters could easily mask the true mass trend. As mentioned above, a larger sample of clusters would be required to investigate this point in detail.

\section{Weak Lensing Results}

In this Section, the results of the analysis of the synthetic weak lensing catalogues are presented, including NFW fits to the data and non-parametric mass maps.

\begin{table*}
\caption{Concentrations, c, and virial radii, $r_{200}$, of clusters from an
  MCMC analysis of simulated weak lensing galaxy catalogues. We consider all
  three simulation prescriptions - cooling and star formation (CSF), cooling
  and star formation with AGN feedback (CSFBH), and dark-matter-only (DM). The
  values shown are for a single projection only (xy). For each value the errors are
  calculated using the 67\% likelihood contours.}
\centering
\begin{tabular}{c | c c c | c c c}
\hline \hline
 Cluster & & c & & & $r_{200}$\,(Mpc) & \\ [0.25ex]
& CSF & CSFBH & DM & CSF & CSFBH & DM \\ [0.5ex]
\hline
C1 & $9.05_{-2.73}^{+3.65}$ & $3.1_{-1.52}^{+1.80}$ & $2.8_{-1.43}^{+1.76}$ & $0.80_{-0.1}^{+0.11}$ & $0.94_{-0.15}^{+0.19}$ & $0.97_{-0.19}^{+0.13}$ \\ [1ex]
C2 &  $10.8_{-1.68}^{+1.7}$ & $9.81_{-1.67}^{+2.19}$ & $5.11_{-0.83}^{+0.81}$ & $1.02_{-0.07}^{+0.07}$ & $1.01_{-0.11}^{+0.09}$ & $0.96_{-0.12}^{+0.12}$\\ [1ex]
C3 & $5.21_{-0.51}^{+0.48}$ & $2.06_{-0.94}^{+0.98}$ & $2.8_{-0.97}^{+1.16}$ & $1.27_{-0.08}^{+0.09}$ & $1.52_{-0.24}^{+0.35}$ & $1.28_{-0.17}^{+0.19}$ \\ [1ex]
C4 &  $8.91_{-1.30}^{+1.61}$ & $6.9_{-1.00}^{+1.01}$ & $5.4_{-0.82}^{+0.87}$ & $1.20_{-0.10}^{+0.10}$ & $1.20_{-0.09}^{+0.11}$ & $1.33_{-0.09}^{+0.10}$\\ [1ex]
C5 & $5.21_{-0.62}^{+0.67}$ & $4.75_{-0.45}^{+0.44}$ & $3.52_{-0.37}^{+0.39}$ & $1.61_{-0.09}^{+0.04}$ & $1.53_{-0.11}^{+0.06}$ & $1.49_{-0.11}^{+0.13}$ \\ [1ex]
\hline
\end{tabular}
\label{table:NFWparams}
\end{table*}

\subsection{NFW profile fits}
{\small COSMOMC} \citep{lewbri} is used to fit NFW profiles \citep{NFW} to the simulated weak lensing data, starting outside the strong lensing regime. The NFW profile can be parameterised with a virial radius $r_{200}$, and a concentration parameter $c$. This allows us to define a scale radius $r_{\rm{s}}=r_{200}/c$. Inside the virial radius, the mass density of the halo equals $200 \rho_{\rm{c}}$, where $\rho_{\rm{c}} = \frac{3 H^{2}(z)}{8 \pi G}$ is the critical density of the universe at the redshift of the halo. We may then write the density profile of the NFW model as,

\begin{equation}
\rho (r) = \frac{\delta_{\rm{c}} \rho_{\rm{c}}}{\left(r/r_{\rm{s}} \right) \left(1 + r/r_{\rm{s}} \right)^{2}}\;,
\end{equation}
 
 where the characteristic overdensity of the halo, $\delta_{\rm{c}}$ can be written,
 
 \begin{equation}
 \delta_{\rm{c}} = \frac{200}{3} \frac{c^3}{\ln \left(1+c \right) - c/\left( 1+ c \right)}\;.
 \end{equation}

Using the results of \cite{BNFW} we can write down the corresponding convergence and shear profiles, which are included in Appendix A.

\begin{figure*}
\centering
\begin{tabular}{cc}
\includegraphics[width=3.5in]{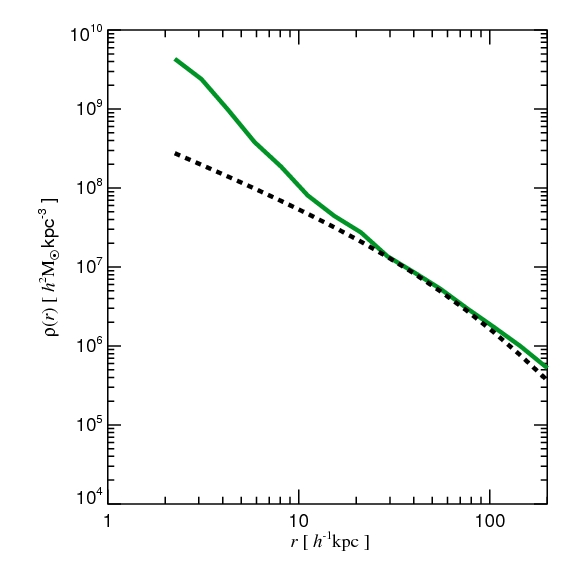}&
\includegraphics[width=3.5in]{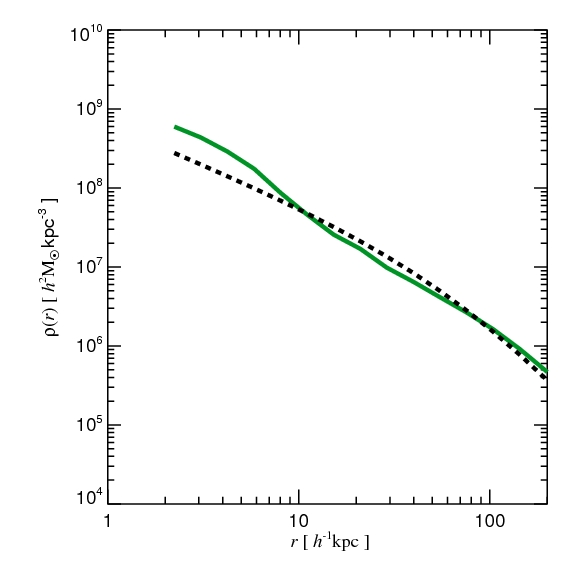}
\end{tabular}
\caption{NFW fits to the total density of cluster C4 simulated both with CSF (left)
  and CSFBH (right). The green solid curve is the total density profile of the cluster and the dashed black curve is the NFW fit. A background galaxy density of 150 arcmin$^{-2}$ is used, and the
  best-fit NFW profiles are computed using an MCMC approach. As expected, we
  see that the NFW profile is a particularly poor fit to the CSF cluster,
  while it fits the CSFBH profile somewhat better.}
\label{NFWfit}
\end{figure*}

The probability distribution for the \emph{observed} galaxy ellipticities, $p_{\epsilon}$, can be obtained from the \emph{intrinisc} ellipticity distribution $p_{\epsilon^{\rm s}}$ (Eq.\,\ref{epdf}) as follows,
\begin{equation}
p_{\epsilon}(\epsilon |g) = p_{\epsilon^{\rm s}} \left(\epsilon^{\rm s} \left( \epsilon |g\right)\right)\left
|{\partial^{2}\epsilon^{\rm s}}\over{\partial \epsilon^{2}}\right | = p_{\epsilon^{\rm s}}(\epsilon^{\rm s}(\epsilon |g))
{(\left | g\right
|^{2}-1)^{2} \over\left |\epsilon g^{*}-1\right |^{4}}\;.
\end{equation}
Using this, we can compute the log-likelihood function from the probability density for each lensed galaxy, $p_{\epsilon}\left(\epsilon_{i}\right)$,
\begin{equation}
\ell_\gamma=-\sum_{i=1}^{N_\gamma}{\ln p_\epsilon(\epsilon_i|g(\bt_i))}\;,
\end{equation}
which can be evaluated numerically for a trial mass model (with corresponding reduced shear field $g(\bt)$) given the lensed ellipticities. The above likelihood calculation is implemented as a module in {\small COSMOMC} \citep{lewbri}, using it as a generic sampler to explore parameter space with the standard Metropolis-Hastings algorithm (\cite{Met}; \cite{Has}). 

\begin{figure*}
\begin{tabular}{lll}
\hspace*{-1.4cm}
\includegraphics[width=2.5in]{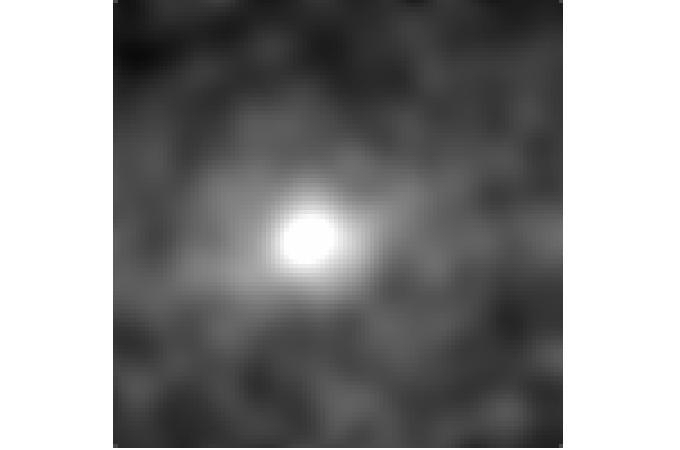}&
\includegraphics[width=2.5in]{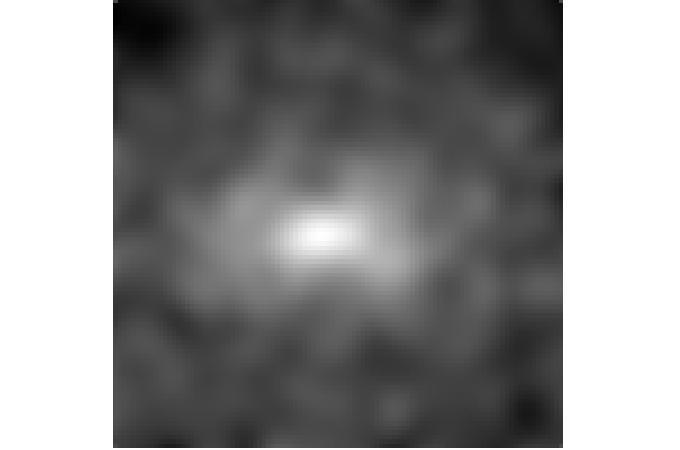}&
\includegraphics[width=2.5in]{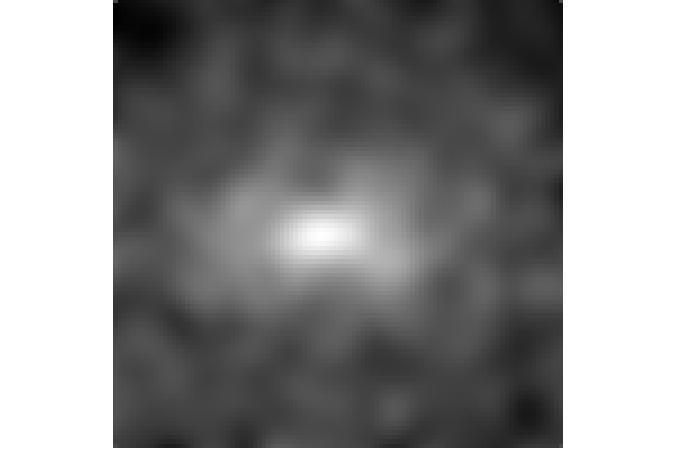}
\end{tabular}
\caption{Non-parametric mass reconstructions of the cluster C5 performed using the algorithm of Seitz \& Schneider (1996). From left to right mass-reconstructions of C5 simulated with CSF, CSFBH and DM are shown, using a background galaxy density of 150 arcmin$^{-2}$ and a smoothing length of 0.38\,arcmin. The size of the box is 0.75 $h^{-1}$\,Mpc on a side. Note that it is almost impossible to distinguish the CSFBH and DM simulations using the mass-reconstruction alone.}
\label{massrecs}
\end{figure*}

The concentrations and virial radii resulting from this analysis are given in Table \ref{table:NFWparams}. 
In clusters C1 and C3, the effect of AGN feedback is to significantly reduce the concentration from the CSF case. In the other clusters, the concentration has still been reduced by the inclusion of AGN feedback, but not as significantly (insofar as the error bars on $c$ for the two types of physics overlap). In contrast, in all the clusters, the concentrations of the DM clusters are only just below those of the CSFBH clusters. The high concentrations observed for the CSF case are due to the very steep inner density profile of CSF clusters - galaxies closer to the cluster centre contribute comparatively more to the likelihood and thus the MCMC algorithm attempts to fit an NFW to this very steep inner density region which leads to inflated concentrations. 

As shown in Fig.\,\ref{NFWfit}, as is to be expected, the NFW profile is a very poor fit to CSF clusters, due to the large central density spike. The NFW is a better fit to the CSFBH clusters but is still unable to reproduce the density profile in the inner regions of the cluster. At larger radii, the CSFBH and DM profiles converge, at a location dependent on the properties of the cluster. For the more massive clusters that would in practice be studied individually with weak gravitational lensing, that the $c$ and $r_{200}$ values are comparable for the CSFBH and DM simulations bodes well for the study of such dark matter profiles using weak lensing. 

In general, we expect the concentration of the halo to rise as halo mass
decreases, and this general trend is observed, with the exception of cluster
C3. As noted earlier, in this cluster several dark matter clumps are merging at the cluster
centre, and this will reduce the concentration measured using a weak-lensing
analysis when fitting a single NFW profile to the entire mass profile.

\subsection{Non-parametric Mass Reconstructions}
The NFW models that we fit to the shear data above are circularly symmetric, and it is clear from 
Fig. \ref{kappamaps} that this assumption of symmetry does not hold for many of the projections.  It is also possible to use the simulated weak lensing data to perform non-parametric mass reconstructions of the clusters, which provides complementary information on the detailed structure of the clusters. Non-parametric mass reconstructions use a statistical analysis of many weakly lensed galaxies to identify over-densities in the mass distribution. In this paper, we perform finite-field reconstructions using the algorithm of \cite{seitzrec}.   

In Fig.\,\ref{massrecs} we show mass reconstructions of C5 simulated with CSF, CSFBH and DM. From the Figure we see that the CSFBH and DM clusters are virtually indistinguishable in the mass reconstructions. The CSF cluster shows a higher central density peak in the reconstruction, as expected.

\section{Discussion and Conclusions}

In this paper we have investigated the impact of baryonic physics, in particular AGN feedback, on the strong-lensing cross-section of a sample of five simulated clusters. We have also briefly considered their weak-lensing properties. 

Using the cluster simulations of \cite{PSS}, the first high-resolution
cosmological simulations to include a self-consistent prescription of BH formation and AGN feedback, we have simulated gravitational lensing by clusters with three different physical prescriptions,

\begin{itemize}

\item Dark-matter only (DM)

\item Cooling and star formation (CSF)

\item The above plus AGN feedback (CSFBH)

\end{itemize}

We have demonstrated that the inclusion of AGN physics can have a large impact on the lensing properties of clusters by both reducing the strong-lensing cross-section relative to CSF physics, and altering the concentrations measured in a weak-lensing analysis. The principle conclusions of this paper are as follows:

\begin{itemize}

\item  Inclusion of baryonic physics can have a dramatic impact on the strong lensing properties of simulated clusters. In CSF simulations we find that the boost in cross-section relative to DM clusters can be significant - our most massive cluster is boosted by a factor of 2-3. The boost is caused by the steepening of the density profile, due to an excess of stars at the cluster centre.

\item We have demonstrated that with the inclusion of AGN feedback, the boost in the cross-section over DM clusters is significantly attenuated relative to those simulations which include only cooling and star formation (CSF). For all but the most massive cluster (C5), where it should be noted that feedback is likely under-predicted, the CSFBH and DM cross-sections are similar. Even for cluster C5, the CSFBH cross-section is $<2$ times greater than the DM cross-section. 

\item The AGN feedback suppresses the build-up of stars and gas in the central
  regions of clusters, ensuring that the CSFBH inner density slope is
  shallower than in CSF haloes - this leads to a reduction in the
  cross-section. The results show some indication for mass dependence of the
  effect - in more massive clusters the CSF cross-section exceeds the
  CSFBH cross-section by more, whilst the ratio of CSFBH to DM is approximately constant. To fully investigate this interesting result, a larger sample of clusters is required.

\item We have shown that baryonic physics can have a measurable effect on weak lensing results. The NFW profile is, as expected, a poor fit to CSF clusters and is not an ideal fit to CSFBH clusters either. Due to the large central density spike of the CSF clusters, measured concentrations can be higher than for the CSFBH and DM clusters. In the mass reconstructions a similar effect is encountered - we see evidence for a prominent central density `spike' in the CSF case.

\end{itemize}

This work has direct implications for the use of clusters in cosmological parameter estimation. Arc abundance is highly sensitive to baryonic physics, making small samples of clusters difficult to use as a tool for precision cosmology. Rather, for small samples, using lensing to probe cluster properties is a much more fruitful endeavour. In this paper we have demonstrated the sensitivity of cluster lensing properties to the baryonic physics of the cluster, and so if we wish to compare observed arc abundances to those in simulations, we must ensure that the baryons are simulated as accurately as possible. This has important consequences for the interpretation of data from large upcoming surveys. The observational prospects for studies of galaxy clusters are bright: for example DES will contain $\sim 20,000$ clusters with mass in excess of $2\times 10^{14}$M$_{\odot}$, up to $\sim 1000$ of which could contain strongly lensed arcs. This will also allow for much larger, homogeneously selected,  samples of arcs to be obtained.

Aside from the giant arc cross-section considered here, as can be seen from Fig.\,\ref{clumag} the DM-only clusters show less substructure than the clusters with baryons, in particular those without AGN feedback. This will influence the appearance of giant arcs that are formed, but also increase the cross-section for the production of smaller separation multiple images by structure outside the critical region of the cluster itself \citep{King07}.

Our key finding is that with the inclusion of AGN feedback, the addition of baryons to dark-matter haloes boosts the cross-section for the production of arcs by less than previously thought. Clusters simulated with AGN feedback have strong lensing cross-sections that are comparable to those simulated with dark matter only. The inclusion of baryonic physics alone cannot be invoked as a means to increase the arc abundance much beyond that predicted in DM simulations. However, baryonic physics is just one part of a larger physical picture, and a realistic comparison with observed arc abundances would require not only simulation of AGN feedback, but also a realistic source population and the inclusion of large scale structure. This is a topic for future work.

\section{acknowledgments}
 JMGM thanks STFC for a postgraduate award, LJK thanks the Royal Society for a
 University Research Fellowship, DS thanks STFC for a postdoctoral fellowship,
 IM thanks the Kavli foundation for a fellowship and AL thanks STFC for a Dorothy Hodgkin
 postgraduate award. Part of the cluster simulations were performed on the Cambridge High Performance Computing Cluster DARWIN in Cambridge (http://www.hpc.cam.ac.uk). We would like to thank Antony Lewis for many helpful discussions, and Volker Springel for reading the manuscript. We would like to thank the referee for helpful suggestions that have greatly improved this manuscript. We would also like to thank Stuart Rankin for his assistance with DARWIN.

\appendix
\section{The Convergence And Shear of The NFW Profile}

The convergence can be written in terms of a dimensionless radial coordinate $x=r/r_{\rm{s}}$ thus,

\begin{equation}
\kappa(x) = \kappa_{k} f(x)\;,
\end{equation}

\noindent where,

\begin{align*}
\displaystyle f\left(x < 1 \right)= \frac{1}{x^{2} -1} \left(1-\frac{2\; \rm{atanh}\;{\sqrt{\frac{1-x}{1+x}}}}{\sqrt{1-x^{2}}} \right) ~~\\
\displaystyle f\left(x =1 \right) = \frac{1}{3} ~~~~~~~~~~~~~~~~~~~~~~~~~~~~~~~~~~~~~~~~ \tag{A2}\\
\displaystyle f\left(x > 1 \right) = \frac{1}{x^{2} -1} \left(1-\frac{2 \arctan {\sqrt{\frac{x-1}{1+x}}}}{\sqrt{x^{2}-1}} \right)~~
\end{align*}

\noindent and

\begin{equation}
\kappa_{k} = \frac{2 r_{s} \delta_{c} \rho_{c}}{\Sigma_{c}}  \tag{A3}
\end{equation}

\noindent The shear is given by,

\begin{equation}
\gamma(x) = \kappa_{k} j(x) \tag{A4}\;,
\end{equation}

\noindent where,

\begin{align*}
\displaystyle j\left(x < 1 \right)= \frac{4 \; \rm{atanh} \; \sqrt{\frac{1-x}{1+x}}}{x^{2}\sqrt{1-x^{2}}} + \frac{2 \ln \left(\frac{x}{2} \right)}{x^{2}} - \frac{1}{x^{2} - 1} \\ 
\displaystyle +  \frac{2 \; \rm{atanh} \; \sqrt{\frac{1-x}{1+x}}}{\left(x^{2}-1 \right) \sqrt{1-x^{2}}} ~~~~~~~~~~~~~~~~~~~~~~ \\
\displaystyle j\left(x =1 \right) = 2 \ln \left(\frac{1}{2} \right) + \frac{5}{3} ~~~~~~~~~~~~~~~~~~~~~~~~~~~~~~\tag{A5}\\
\displaystyle j\left(x > 1 \right) = \frac{4 \arctan \sqrt{\frac{x-1}{1+x}}}{x^{2}\sqrt{x^{2}-1}} + \frac{2 \ln \left(\frac{x}{2} \right)}{x^{2}} - \frac{1}{x^{2} - 1} \\
\displaystyle +   \frac{2 \arctan \sqrt{\frac{x-1}{1+x}}}{\left(x^{2}-1 \right)^{\frac{3}{2}}}~~~~~~~~~~~~~~~~~~~~~~~~~ \\
\end{align*}

\appendix
\label{lastpage}

\end{document}